\documentclass{article}
\usepackage[T1]{fontenc} 
\usepackage[utf8]{inputenc} 
\usepackage{ismir,amsmath,cite,url}
\usepackage{subcaption}
\usepackage{graphicx}
\usepackage{color}
\usepackage{multirow}
\usepackage{makecell}

\usepackage{lineno}

\title{On Perceived Emotion in Expressive Piano Performance:\\Further Experimental Evidence for the Relevance of Mid-level Perceptual Features}

\multauthor
{Shreyan Chowdhury$^1$ \hspace{0.5cm} Gerhard Widmer$^{1,2}$} { 
$^1$Institute of Computational Perception, Johannes Kepler University Linz, Austria\\
$^2$LIT AI Lab, Linz Institute of Technology, Austria\\
{\tt\small {firstname.lastname}@jku.at}
}
\def\authorname{Shreyan Chowdhury, Gerhard Widmer}

\def\authorname{S. Chowdhury, and G. Widmer}

\begin{document}

\maketitle
\begin{abstract}
Despite recent advances in audio content-based music emotion recognition, a question that remains to be explored is whether an algorithm can reliably discern emotional or expressive qualities between different performances of the same piece. In the present work, we analyze several sets of features on their effectiveness in predicting arousal and valence of six different performances (by six famous pianists) of Bach's Well-Tempered Clavier Book 1. These features include low-level acoustic features, score-based features, features extracted using a pre-trained emotion model, and Mid-level perceptual features. We compare their predictive power by evaluating them on several experiments designed to test performance-wise or piece-wise variations of emotion. We find that Mid-level features show significant contribution in performance-wise variation of both arousal and valence -- even better than the pre-trained emotion model. Our findings add to the evidence of Mid-level perceptual features being an important representation of musical attributes for several tasks -- specifically, in this case, for capturing the expressive aspects of music that manifest as perceived emotion of a musical performance.
\end{abstract}
\section{Introduction}\label{sec:introduction}



A musical performance, particularly in the Western music tradition, is not merely a literal acoustic rendering of a notated piece or composition. Rather, the piece is transformed by the performer's own expressive performance choices, relating to such dimensions as the choice of tempo, expressive tempo and timing variations, dynamics, articulation, and so on. The emotional effect of a performance on a listener can be a consequence both of the composition itself, with its musical properties and structures, and of the performance, the way the piece was played. 
In fact, it has been convincingly demonstrated
\cite{Gabrielsson1996Emotional,akkermans2019decoding}
that performers are capable of communicating, with high accuracy, intended emotional qualities by their playing.

The analysis of emotion in music recordings has a long history in Music Information Retrieval, with many works addressing content-based emotion regression and classification typically using low-level or hand-crafted audio and musical features~\cite{panda2020features, soleymani2014emotional, kim2010music, yang2012machine} or using deep learning based methods~\cite{orjesek2019dnn, er2019music, he2020multi}. 
However, there has been little research on the more subtle problem of identifying emotional aspects that are due to the actual \textit{performance}, and even less on models that can automatically recognize this from \textit{audio} recordings. On the latter problem -- the one to be addressed in this paper -- the most directly relevant prior work we are aware of is \cite{grekow2018musical}, where 324 6-second audio snippets of different genres (classical, jazz, blues, metal, etc.) were annotated in terms of perceived emotion (valence and arousal), and various regressors were trained to predict these two dimensions from a set of standard audio features. The regression models were then used to predict valence-arousal trajectories over 5 different recordings of 4 Chopin pieces, but no ground truth in terms of human emotion annotations was collected. The relevance of the model predictions was evaluated only indirectly, by comparing similarity scores between predicted profiles with overall performance similarity ratings by three human listeners,
which showed some non-negligible correlations.




In a recent focused study~\cite{battcock2019acoustically}, Battcock \& Schutz (referred to as ``B\&S'' henceforth) investigate how three specific score-based cues (Mode, Pitch Height, and Attack Rate\footnote{Actually, attack rate as computed by B\&S is also informed by the average tempo of the performance; thus, it is not strictly a score-only feature.}) work together to convey emotion in J.S.Bach's preludes and fugues collected in his \textit{Well-tempered Clavier (WTC)}. They used recordings of the complete WTC Book 1 (48 pieces) of one famous pianist (Friedrich Gulda) as stimuli for human listeners to rate each performance on perceived arousal and valence. Their findings suggest that within this set of performances, arousal is significantly correlated with attack rate and valence is affected by both the attack rate and the mode. However, that study was based on only one set of performances, making it impossible to decide whether the human emotion ratings used as ground truth really reflect aspects of the compositions themselves, or whether they were also (or even predominantly) affected by the specific (and, in some cases, rather unconventional) way in Friedrich Gulda plays the pieces -- that is, whether the emotion ratings reflect piece or performance aspects.

The purpose of the present paper is to try to disentangle the possible
contributions and roles of different features in capturing composer-(piece-)specific
and performer-(recording-)specific aspects. To this end, we collected
human ratings of perceived valence and arousal in six complete sets of recordings
of WTC Book 1, and then performed a systematic study with feature sets derived from various levels of musical abstraction, including some extracted by pre-trained deep neural networks.



\section{Data Collection}\label{sec:data}

\subsection{Pieces and Recordings}

J.S.Bach's \textit{Well-tempered Clavier (WTC)} is ideally suited for systematic and controlled studies of this kind, as it comprises a stylistically coherent set of keyboard pieces from a particular period, evenly distributed over all keys and major/minor modes, with a pair of two pieces (a prelude, followed by a fugue) in each of the 24 possible keys, for a total of 48 pieces. Each piece has its own distinctive musical character, and despite being written in a rather strict style and not meant to be played in `romantic' ways, the music offers pianists (or pianists take) lots of liberties in ornamentation, but also overall performance parameters (e.g., tempo and articulation). For example, there are pieces in our set of recordings that one pianist takes more than twice~(!) as fast as another.

For a broad set of diverse performances, we selected six recordings of the complete WTC Book 1, by six famous and highly respected pianists, all of whom can be considered Bach specialists to various degrees. The recordings are listed in Table \ref{tab:recordings}.

\begin{table}[t]
\small
  \begin{tabular}{l|l|l}
  Pianist & Recording & Year \\
  \hline
  Glenn Gould & Sony 88725412692 & 1962-1965 \\
  Friedrich Gulda & MPS 0300650MSW & 1972 \\
  Angela Hewitt & Hyperion 44291/4 & 1997-1999 \\
  Sviatoslav Richter & RCA 82876623152 & 1970 \\
  Andr\'as Schiff & ECM 4764827 & 2011 \\
  Rosalyn Tureck & DG 4633052 & 1952-1953 
  \end{tabular}%
  \caption{\small Pianists and recordings.}
    \label{tab:recordings}
\end{table}

\subsection{Emotion Annotations and Pre-processing}
In accordance with B\&S, we will only use the first 8 bars of each recording for the annotation process and our experiments. These were cut out manually. The participants of our annotation exercise were students of a course at a university, without a specifically musical background. Each participant heard a subset of the recordings (all 48 pieces as played by one pianist) and was asked to rate the valence on a scale of -5 to +5 (11 levels) and the arousal on a scale of 0 to 100 (increments of 10; a total of 11 levels). They could listen to a recording as many times as they liked. Each recording was rated by 29 participants. In total, we collected 8,352 valence-arousal annotation pairs.

\begin{figure}[h]
    \centering
    \includegraphics[clip,trim=1cm 0cm 2cm 0cm, width=\columnwidth]{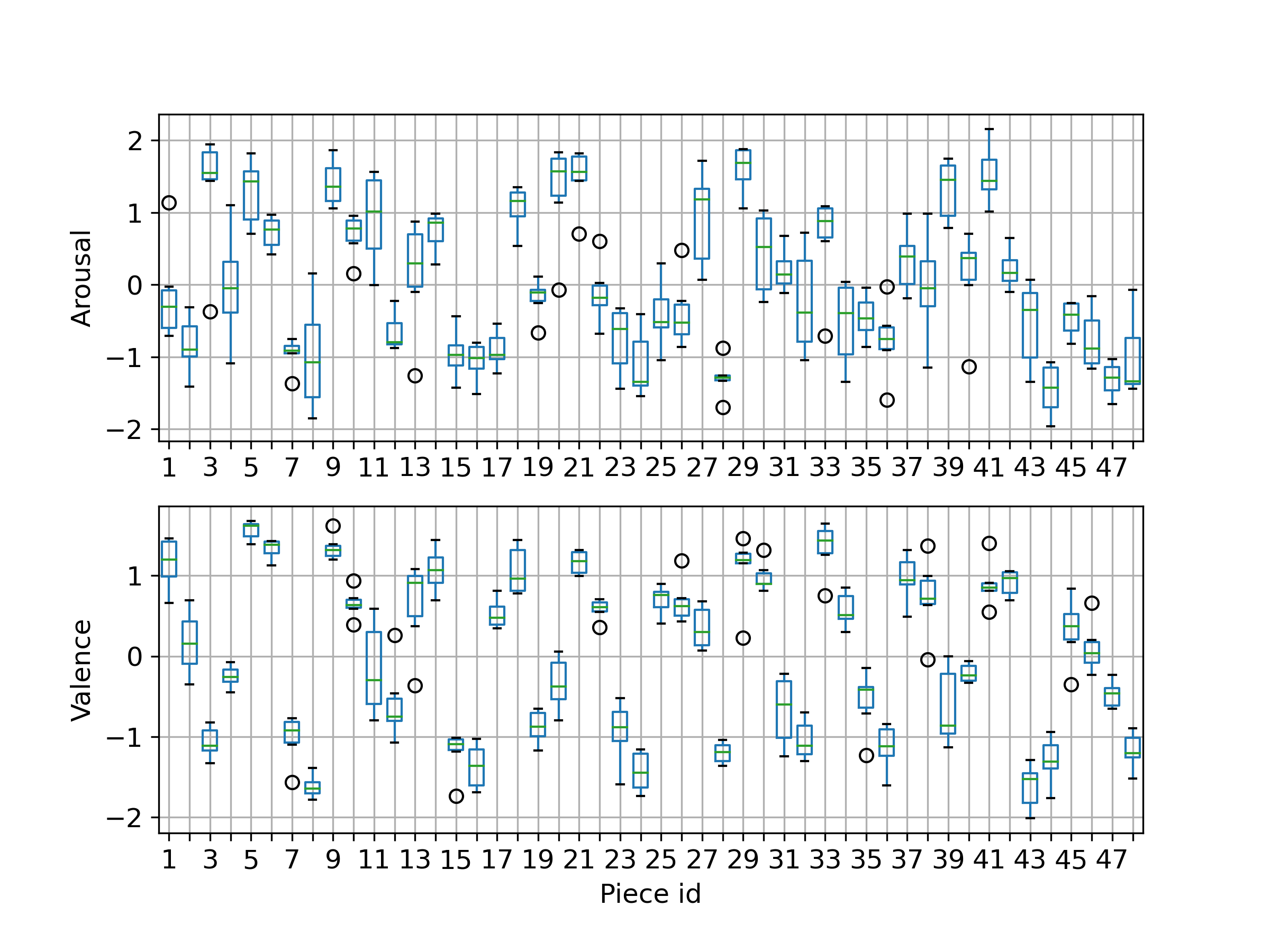}
    \caption{\small Distribution of emotion ratings across pianists for every piece.}
    \label{fig:boxplots}
\end{figure}

For the purposes of this paper, we take the mean arousal and mean valence ratings for each recording, and these values serve as our ground-truth values for all following experiments. The distributions (over the 6 performances) of these mean ratings for each piece are summarised in Figure~\ref{fig:boxplots}. 

\section{Feature Sets}\label{sec:features}
In this section, we briefly describe the four feature sets that we use to model arousal and valence. 

\subsection{Low-level Features}
These consist of hand-crafted musical features (such as onset rate, tempo, pitch salience) as well as generic audio descriptors (such as spectral centroid, loudness). Taken together, they reflect several musical characteristics such as tone colour, dynamics, and rhythm. A brief description of all low-level features that we use is given in Table~\ref{tbl:ll_features}. We use Essentia~\cite{bogdanov2013essentia} and Librosa~\cite{mcfee2015librosa} for extracting these. The audio is sampled at 44.1kHz and the spectra computed (when required) with a frame size of 1024 samples and a hop size of 512 samples. Each feature is aggregated over the entire duration of an audio clip by computing the mean and standard deviation over all the frames of the clip (a `clip' being an 8 bar initial segment from a recording).
\begin{table}[h]
\small
\resizebox{\columnwidth}{!}{%
	\begin{tabular}{>{\centering}m{2.3cm}|m{5.5cm}}
      \hline
      Dissonance & Total harmonic dissonance computed from pairwise dissonance of all spectral peaks. \\
      \hline
      Dynamic Complexity & The average absolute deviation from the global loudness level estimate in dB. \\
      \hline
      Loudness & Mean loudness of the signal computed from the signal amplitude. \\
      \hline
      Onset Rate & Number of onsets (note beginnings or transients) per second. \\
      \hline
      Pitch Salience & A measure of tone sensation, computed from the harmonic content of the signal. \\
      \hline
      Spectral Centroid & The weighted mean frequency in the signal, with frequency magnitudes as the weights.\\
      \hline
      Spectral Flatness & A measure to quantify how much noise-like a sound is, as
      opposed to being tone-like. \\
      \hline
      Spectral Bandwidth & The second order bandwidth of the spectrum. \\
      \hline
      Spectral Rolloff & The frequency under which 85\% of the total energy of the spectrum is contained. \\
      \hline
      Spectral Complexity & The number of peaks in the input spectrum. \\
      \hline
      Tempo (BPM) & Tempo estimate from audio in beats per minute. \\
      \hline
	\end{tabular}}
	\caption{\small Low-level Features}
	\label{tbl:ll_features}
\end{table}

\subsection{Score Features}\label{sec:feat_description_score}
The following set of features was computed directly from the musical score (i.e., sheet music) of the pieces instead of the audio files. The unit of score time, ``beat'', is defined by the time signature of the piece (e.g., 4/4 means that there are 4 beats of duration 1 quarter in a bar). The score information and the audio files were linked using automatic score-to-performance alignment. Table~\ref{tbl:score_features} describes the score features in detail. 

\begin{table}[h]
\small
\resizebox{\columnwidth}{!}{%
	\begin{tabular}{>{\centering}m{1.8cm}|m{6cm}}
      \hline
      Inter Onset Interval & The time interval between consecutive notes per beat.\\
      \hline
      Duration & Two features describing the empirical mean and standard deviation of the notated duration per beat in the snippet. \\
      \hline
      Onset Density & The number of note onsets per beat. A chord constitutes a single onset. \\
      \hline
      Pitch Density & The number of unique notes per beat. \\
      \hline
      Mode & Binary feature denoting major/minor modality, computed using the Krumhansl-Schmuckler key finding algorithm~\cite{krumhansl2001cognitive} (to reflect the fact that the dominant key over the segment may be different from the given key signature).\\
      \hline
      Key Strength & This feature represents how much does the tonality by the "Mode" feature fit the snippet. \\
      \hline
	\end{tabular}}
	\caption{\small Score Features}
	\label{tbl:score_features}
\end{table}

\subsection{Mid-level Features}
Mid-level features, described in~\cite{Aljanaki2018Midlevel}, are perceptual musical features that are intuitively understandable to the average listener. They seem well-suited to bridge the ``semantic gap" between low-level audio features and high-level descriptors such as emotion and have been shown to be useful in explainable music emotion recognition~\cite{Chowdhury2019}. We learn these features from the Mid-level Dataset~\cite{Aljanaki2018Midlevel} using a receptive-field regularised residual neural network (RF-ResNet) model~\cite{koutini2019receptive}. Since we intend to use this model to extract features from solo piano recordings (a genre that is not covered by the original training data), we use a domain-adaptive training approach as described in~\cite{chowdhury2021towards}. We use an input audio length of 30 seconds, padded or cropped as required. As these features cannot be strictly defined, Table~\ref{tbl:aljanakifeatures} lists a rough description adapted from the questions in~\cite{Aljanaki2018Midlevel} that were shown to the annotators of the dataset to help them rate the audio clips.
\begin{table}
\small
\resizebox{\columnwidth}{!}{%
	\begin{tabular}{>{\centering}m{1.9cm}|m{5.9cm}}
      \hline
      Melodiousness & How singable is this music? \\
      \hline
      Articulation & Overall impression of articulation in terms of staccato or legato playing style. Higher means more staccato. \\
      \hline
      Rhythmic Stability & How easy is it to march-along with the music? \\
      \hline
      Rhythmic Complexity & How difficult is it to follow the music by tapping? Rhythmic layers and different meters correlate with higher complexity. \\
      \hline
      Dissonance & Noisier timbre or presence of dissonant intervals (tritones, seconds, etc.) \\
      \hline
      Tonal Stability & How clear or apparent the tonic and key are. \\
      \hline
      Minorness & Relates to the perceived tonality. More ``minor-sounding" music will have higher minorness. \\
      \hline
	\end{tabular}
	}
	\caption{\small Mid-level Features
}
	\label{tbl:aljanakifeatures}
\end{table}

\subsection{DEAMResNet Emotion Features}\label{sec:emo_model}
To compare the mid-level features with another deep neural network based feature extractor, we train a model with the same architecture (RF-ResNet) and training strategy on the DEAM dataset~\cite{aljanaki2017developing} to predict arousal and valence from spectrogram inputs. Since this model is trained to predict arousal and valence, it is expected to learn representations suitable for this task. As with the mid-level model, we perform domain adaptation while training this model also.

Features are extracted from the penultimate layer of the model, which gives us 512 features. Since these are too many features to use for our dataset containing only 288 data points, we perform dimensionality reduction using PCA (Principal Component Analysis), to obtain 9 components explaining at least 98\% of the variance. These 9 features are named as \texttt{pca\_x} with \texttt{x} being the principal component number.

\section{Feature Evaluation Experiments}\label{sec:feat_exp}
In this section, we evaluate the four feature sets. The aim of this section is to answer the following questions:
\begin{enumerate}
  \item How well can each feature set fit the arousal and valence ratings? How do these feature sets compare to the ones used by B\&S? (Sections~\ref{sec:bs_eval} and~\ref{sec:our_data})
  \item In each feature set, which features are the most important? (Section~\ref{sec:feat_importance})
  \item Which feature set best explains variation of arousal and valence \textit{between pieces}? (Section~\ref{sec:piece_variation})
  \item Which feature set best explains variation of arousal and valence \textit{between different performances of the same piece}?
  (Section~\ref{sec:performance_variation})
\end{enumerate}
We use ordinary least squares fitting on the dataset in question and calculate the regression metrics. 
The metrics we report are adjusted coefficient of determination ($\tilde{R}^2$), root mean squared error between true and predicted values (RMSE), and Pearson's correlation coefficient between true and predicted values (Corr). 



\subsection{Evaluation on B\&S Data}\label{sec:bs_eval}

As a starting point, we take the data used by B\&S in Experiment 3 of their paper -- Gulda's performances rated on valence and arousal
We perform regression with our feature sets and compare with the values obtained by B\&S using their features Attack Rate, Pitch Height, and Mode. The results are summarised in Table~\ref{tbl:gulda_bs}. 

\begin{table}[h]

\small
\resizebox{\columnwidth}{!}{%
  \begin{tabular}{l|r|r|r|r|r|r}
    \multirow{2}{*}{} &
      \multicolumn{3}{c|}{Arousal} &
      \multicolumn{3}{c}{Valence} \\
    
    & $\tilde{R}^2$ & RMSE & Corr & $\tilde{R}^2$ & RMSE & Corr  \\
    \hline
    Mid-level & 0.84 & 0.36 & 0.93 & \textbf{0.79} & 0.42 & 0.91  \\
    DEAMResNet & \textbf{0.91} & 0.27 & 0.96 & 0.69 & 0.50 & 0.86  \\
    Low-level & 0.86 & 0.29 & 0.96 & 0.67 & 0.45 & 0.89  \\
    Score & 0.31 & 0.74 & 0.67 & 0.61 & 0.55 & 0.83  \\
    B\&S (exp 3) & 0.48 & - & - & 0.75 & - & -  \\
  \end{tabular}%
  }

    


\caption{\small Regression on Gulda data from B\&S~\cite{battcock2019acoustically}.
}
\label{tbl:gulda_bs}
\end{table}

We can see that all three audio-based features perform considerably well for both arousal and valence to motivate further analysis.


\subsection{Evaluation on Our Dataset}\label{sec:our_data}

Next, we perform regression on our complete dataset (comprising of 288 unique recordings -- 48 pieces $\times$ 6 pianists). The results summary can be seen in Table~\ref{tbl:our_data_regress}. Here again, we observe that while DEAMResNet Emotion features perform best on arousal and Score features perform best on valence, Mid-level features show a balanced performance across both the emotion dimensions. 

To evaluate generalizability, we perform cross-validation with three different kinds of splits -- piece-wise (all 6 performances of a piece are test samples in a fold, for a total of 48 folds), pianist-wise (all 48 pieces of a pianist are test samples in a fold, for a total of 6 folds), and leave-one-out (one recording is the test sample per fold, for a total of 288 folds). This is summarized in Table~\ref{tbl:our_data_cv}.

\begin{table}[h]
\begin{subtable}{\columnwidth}
\small
\resizebox{\columnwidth}{!}{%
  \begin{tabular}{l|r|r|r|r|r|r}
    \multirow{2}{*}{} &
      \multicolumn{3}{c|}{Arousal} &
      \multicolumn{3}{c}{Valence} \\
    
    Feature Set & $\tilde{R}^2$ & RMSE & Corr & $\tilde{R}^2$ & RMSE & Corr  \\
    \hline
    Mid-level & 0.68 & 0.56 & 0.83 & 0.63 & 0.60 & 0.80  \\
    DEAMResNet & \textbf{0.70} & 0.54 & 0.84 & 0.42 & 0.72 & 0.69  \\
    Low-level & 0.62 & 0.59 & 0.81 & 0.41 & 0.74 & 0.67  \\
    Score & 0.41 & 0.75 & 0.65 & \textbf{0.75} & 0.49 & 0.87  \\

  \end{tabular}%
  }
  
  \caption{Regression metrics with our data}
  \label{tbl:our_data_regress}
\end{subtable} 
\begin{subtable}[h]{\columnwidth}
\medskip 
\small
\resizebox{\columnwidth}{!}{%
  \begin{tabular}{l|r|r|r|r|r|r}
    \multirow{2}{*}{} &
      \multicolumn{2}{c|}{Piece-wise} &
      \multicolumn{2}{c|}{Pianist-wise} &
      \multicolumn{2}{c}{LOO} \\
    
    Feature Set & A & V & A & V & A & V  \\
    \hline
    Mid-level & \textbf{0.68} & 0.63 & \textbf{0.68} & 0.64 & \textbf{0.69} & 0.65  \\
    DEAMResNet & 0.67 & 0.37 & 0.61 & 0.41 & 0.68 & 0.43  \\
    Low-level & 0.54 & 0.20 & $-$0.11 & $-$0.05 & 0.57 & 0.30  \\
    Score & 0.08 & \textbf{0.67} & 0.39 & \textbf{0.75} & 0.37 & \textbf{0.74}  \\

  \end{tabular}%
  }
  \caption{$\tilde{R}^2$ for different cross-validation splits. A: Arousal, V: Valence, LOO: Leave-One-Out}
  \label{tbl:our_data_cv}
 \end{subtable}
 \caption{\small Evaluation results on our data}
\end{table}

We see that Mid-level features show good generalization for arousal and are robust to different kinds of splits. They also show balanced performance between arousal and valence for all splits. The good performance of the Score features on the valence dimension (V), here and in the previous experiment, is mostly due to the \textit{Mode} feature; there is a substantial correlation in the annotations between major/minor mode and positive/negative valence.


    

        
        

\begin{figure*}[!ht]
    \centering
    \begin{subfigure}[b]{0.475\textwidth}
        \centering
        \includegraphics[width=\textwidth]{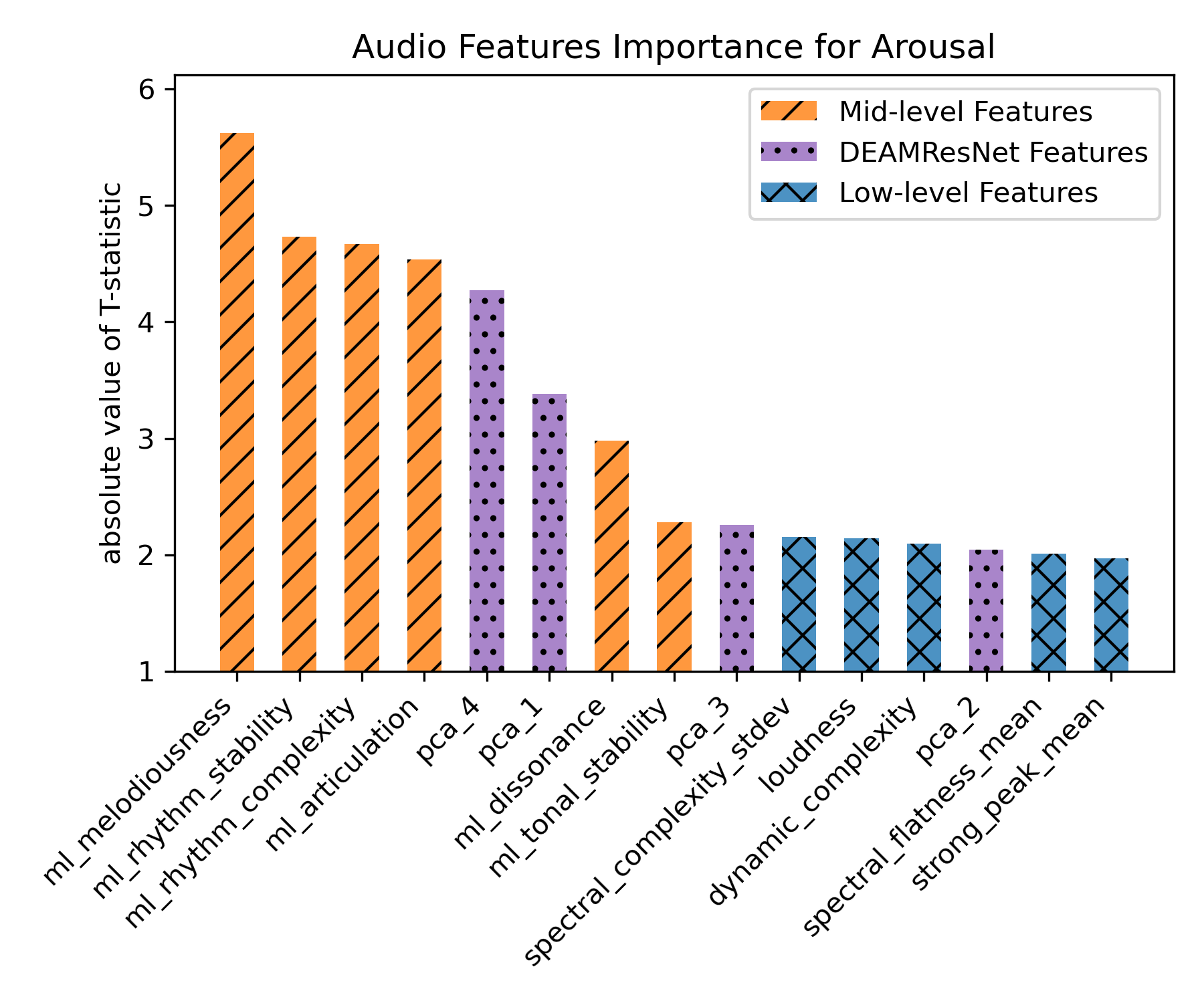}
        \label{fig:audio_features_arousal}
    \end{subfigure}
    \hskip 20pt
    \begin{subfigure}[b]{0.47\textwidth}  
        \centering 
        \includegraphics[width=\textwidth]{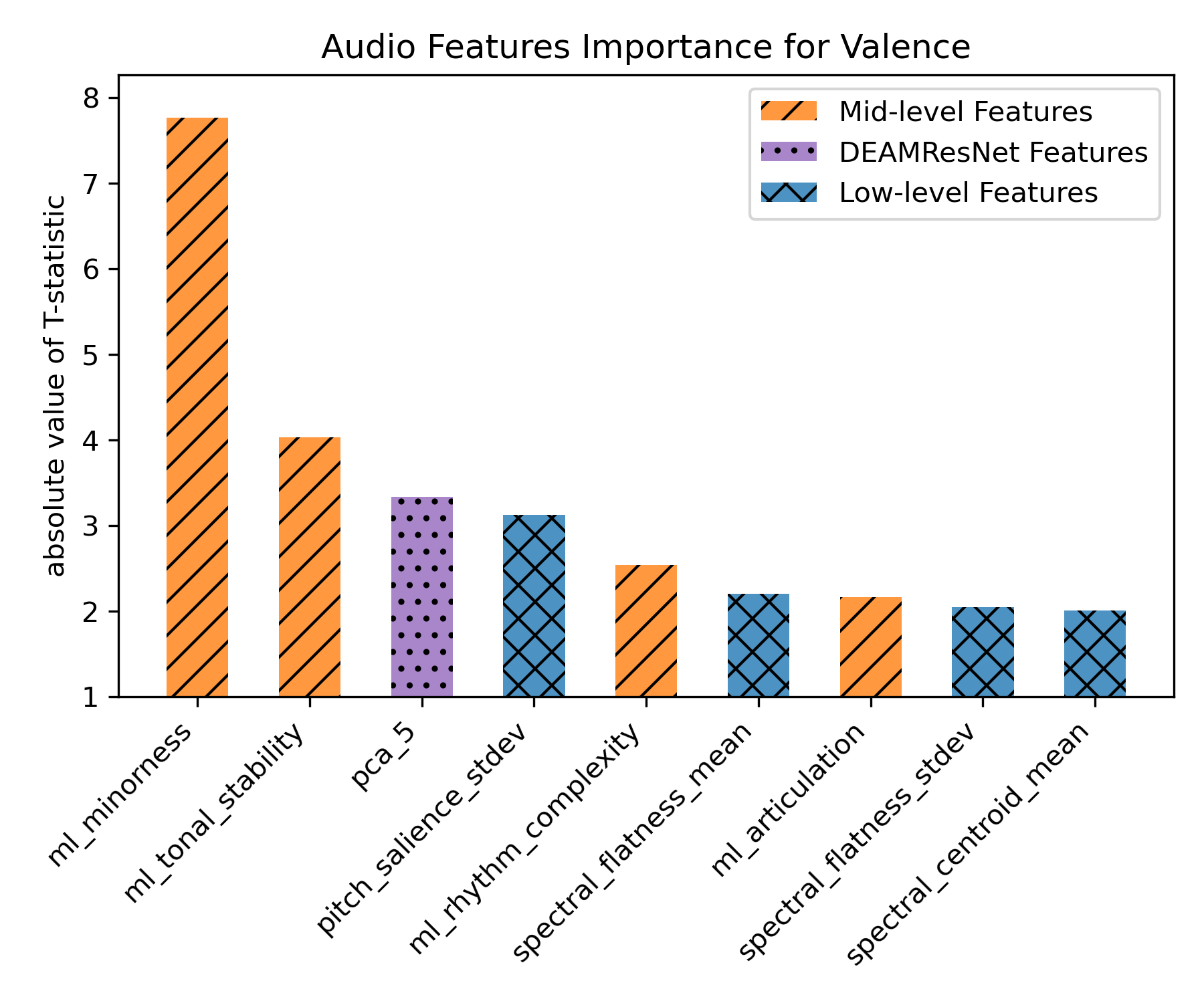}
        \label{fig:audio_features_valence}
    \end{subfigure}
    \vskip -20pt
    \caption{\small Feature importance for audio features using T-statistic. Only features with p<0.05 are shown.
    }
    \label{fig:features_tval}
\end{figure*}

\subsection{Feature Importance within Feature Sets}\label{sec:feat_importance}
We use the absolute value of the t-statistic of a feature as the importance measure. T-statistic is defined as the estimated weight scaled with its standard error. We focus on the audio-based feature sets here, as in most realistic applications scenarios, the score information will not be available (and, being constant across different performances, will not be able to distinguish performance aspects).
We perform a regression using all audio-based features (numbering 39 in total) and compare the t-values in Figure~\ref{fig:features_tval}.

We see that the top-4 and top-2 features in arousal and valence, respectively, are Mid-level features. These features also make obvious musical sense -- modality is often correlated with valence (positive or negative emotional quality), and tempo, rhythm, and articulation with arousal (intensity or energy of the emotion). 

\subsection{Explaining Piece-wise Variation}\label{sec:piece_variation}
We observe from the annotation data (see Figure~\ref{fig:boxplots}) that the distribution of emotion ratings of each piece is distinct (here, the 6 performances for each piece form the ``distribution'' of the piece). In essence, the mean value of arousal or valence depends on the piece in question. Therefore, to take into account this factor of variation, we use linear mixed models~\cite{nlme} to model arousal and valence.

In this linear mixed effect model, the piece id is considered as a ``random effect'' intercept, which models part of the residual remaining unexplained by the features we are evaluating (``fixed effects''). A feature set that models piece-wise variation better than another set would naturally have a lesser residual variation to be explained by the random effect. We therefore look at which feature set has the least fraction of residual variance explained by the random effect of piece id, defined as:

\begin{equation}
    \textrm{E}_{\textrm{random}}=\frac{\textrm{Var}_{\textrm{random}}}{\textrm{Var}_{\textrm{random}} + \textrm{Var}_{\textrm{residual}}}
\end{equation}

where $\textrm{Var}_{\textrm{random}}$ is the variance of the random effect intercept and $\textrm{Var}_{\textrm{residual}}$ is the variance of the residual that remains after mixed effects modeling.

\begin{table}[h]
\small
\centering
  \begin{tabular}{l|r|r}
    
    Feature Set & Arousal & Valence  \\
    \hline
    Mid-level & 0.50 & 0.86  \\
    DEAMResNet & \textbf{0.47} & 0.89  \\
    Low-level & 0.66 & 0.90  \\
    Score & 0.63 & \textbf{0.68} \\

  \end{tabular}%
  \caption{\small Fraction of residual variance explained by the random effect of ``piece id''.}
  \label{tbl:piece_variation}
\end{table}

We see from Table~\ref{tbl:piece_variation} that the DEAMResNet emotion features best explain piece-wise variation in arousal, followed closely by Mid-level features. For valence, the performance of all three audio-based features are close, with Mid-level features performing the best, however, score features outperform them with a large margin. This is again due to the relationship between mode and valence, and mode covarying tightly with the piece ids.

\subsection{Explaining Performance-wise Variation}\label{sec:performance_variation}

Evaluation of performance-wise variation modelling cannot be done with the mixed effects approach as in the previous section because the means (of arousal or valence across all pieces) are nearly identical for each pianist. 

Therefore, we look at one piece at a time and compute the fraction of variance unexplained (FVU) and Pearsons's correlation coefficient (Corr) between predicted and true values across performances for each such test piece. This is done as leave-one-piece-out cross-validation, and aggregated by taking the means. The p-values of the correlation coefficients are counted and we report the percentage of pieces for which $p<0.1$. With only 6 performances per piece, a significance level of $p<0.05$ is obtained for only a handful of pieces. Since score-features based predictions are exactly equal for all performances of a piece, these metrics are not meaningful, and hence the Score feature set is not included here.

\begin{table}[h]
\centering
\resizebox{\columnwidth}{!}{%
  \begin{tabular}{l|r|r|r|r}
    \multirow{2}{*}{} &
      \multicolumn{2}{c|}{Arousal} &
      \multicolumn{2}{c}{Valence} \\
    
    Feature Set & FVU & Corr (p<0.1) & FVU & Corr (p<0.1) \\
    \hline
    Mid-level & \textbf{0.31} & \textbf{0.58}  (47.9\%) & \textbf{0.36} & 0.42 (27.0\%) \\
    DEAMResNet & 0.32 & 0.54 (43.8\%) & 0.61 & \textbf{0.47} (37.5\%)  \\
    Low-level & 0.43 & 0.56 (54.2\%) & 0.75 & 0.38 (22.9\%)  \\

  \end{tabular}%
  }
  \caption{\small Evaluation metrics for performance-wise variation. FVU: Fraction of Variance Unexplained. Corr: Pearson's correlation coefficient.}
\end{table}

\begin{figure}[h]
    \centering
    \includegraphics[width=\columnwidth]{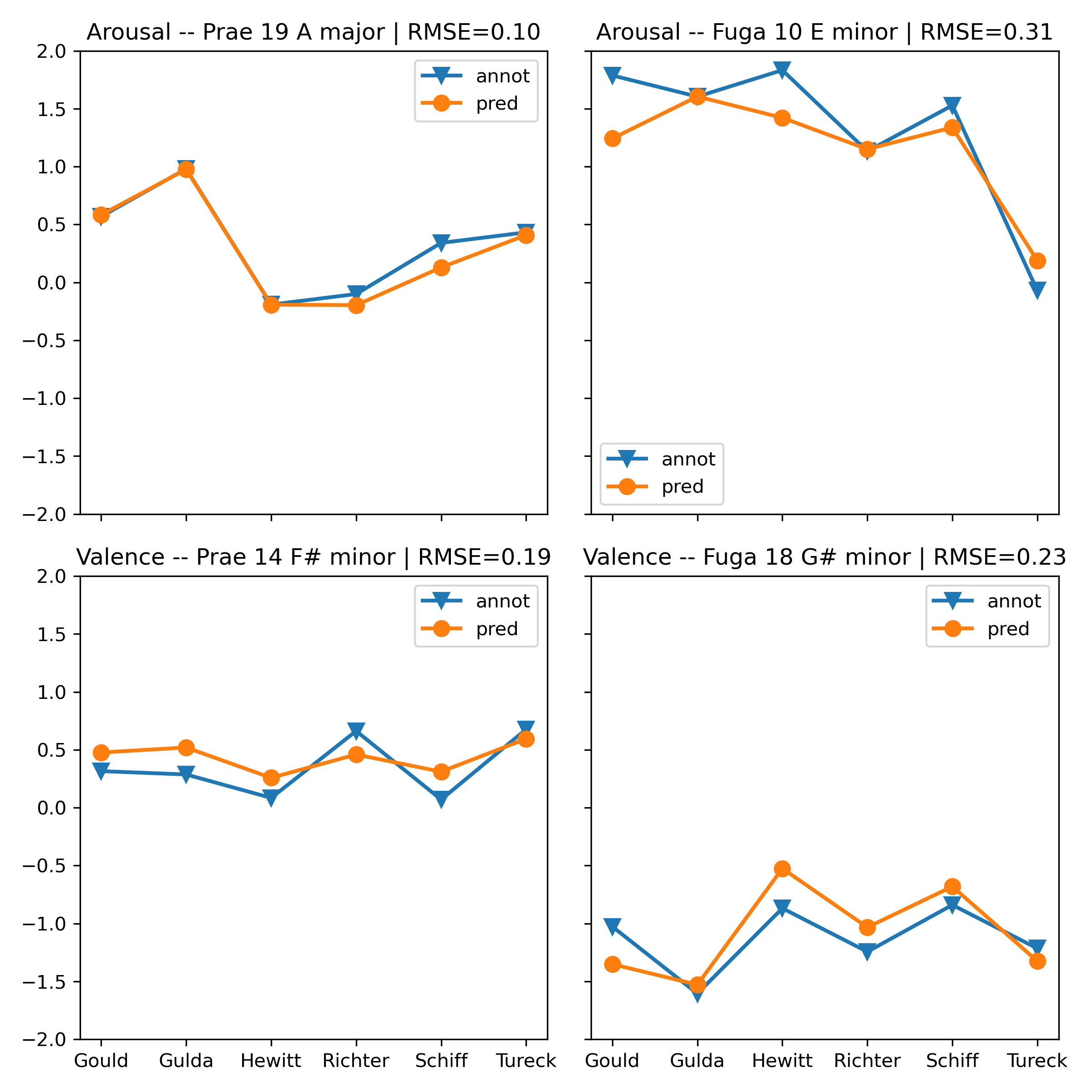}
    \caption{\small Some example pieces with high emotion variability between performances which are modeled particularly well using mid-level features.}
    \label{fig:exhibit}
\end{figure}

Again, Mid-level features come out at the top in most measures. To illustrate the modelling of performance-wise variation, we select a few example pieces that have a high variation of emotion between performances and plot them together with the predicted values using mid-level features in Figure~\ref{fig:exhibit}. The predicted emotion dimensions follow the ratings closely, even for performances that deviate from the average (e.g. the arousals of Gulda's performance of Prelude in A major and Tureck's performance of Fugue in E minor.)

\section{Probing Further}

We now describe two additional experiments designed to further probe the predictive power of our feature sets.

\subsection{Predicting Emotion of Outlier Performances}
\begin{figure}[!h]
    \centering
    \includegraphics[clip,trim=0.4cm 0cm 0.2cm 0cm, width=\columnwidth]{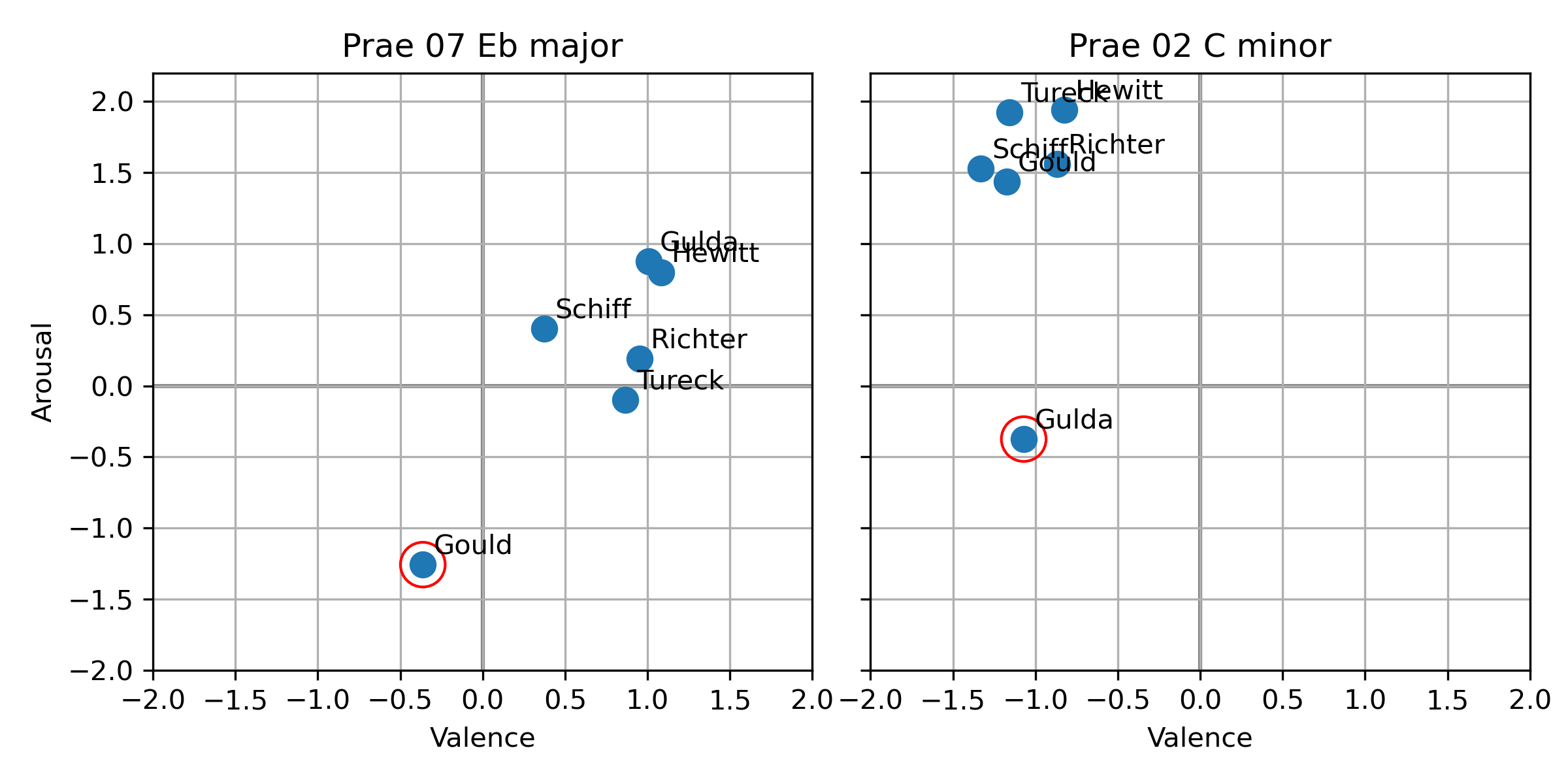}
    \caption{\small Two examples of outlier performances: Prelude \# 7 in Db major, outlier is Gould (left); Prelude \#2 in C minor (Gulda, right).}
    \label{fig:outlier_av}
\end{figure}

Figure~\ref{fig:outlier_av} shows two examples of pieces where one performance has a vastly different emotional character than the others -- in the first example, Gould even produces a negative valence effect (mostly through tempo and articulation) in the E-flat major prelude, which the others play in a much more flowing fashion. A challenge for any model would thus be to predict the emotion of such idiosyncratic performances, not having seen them during training.

\begin{figure}[h]
    \centering
    \includegraphics[width=\columnwidth]{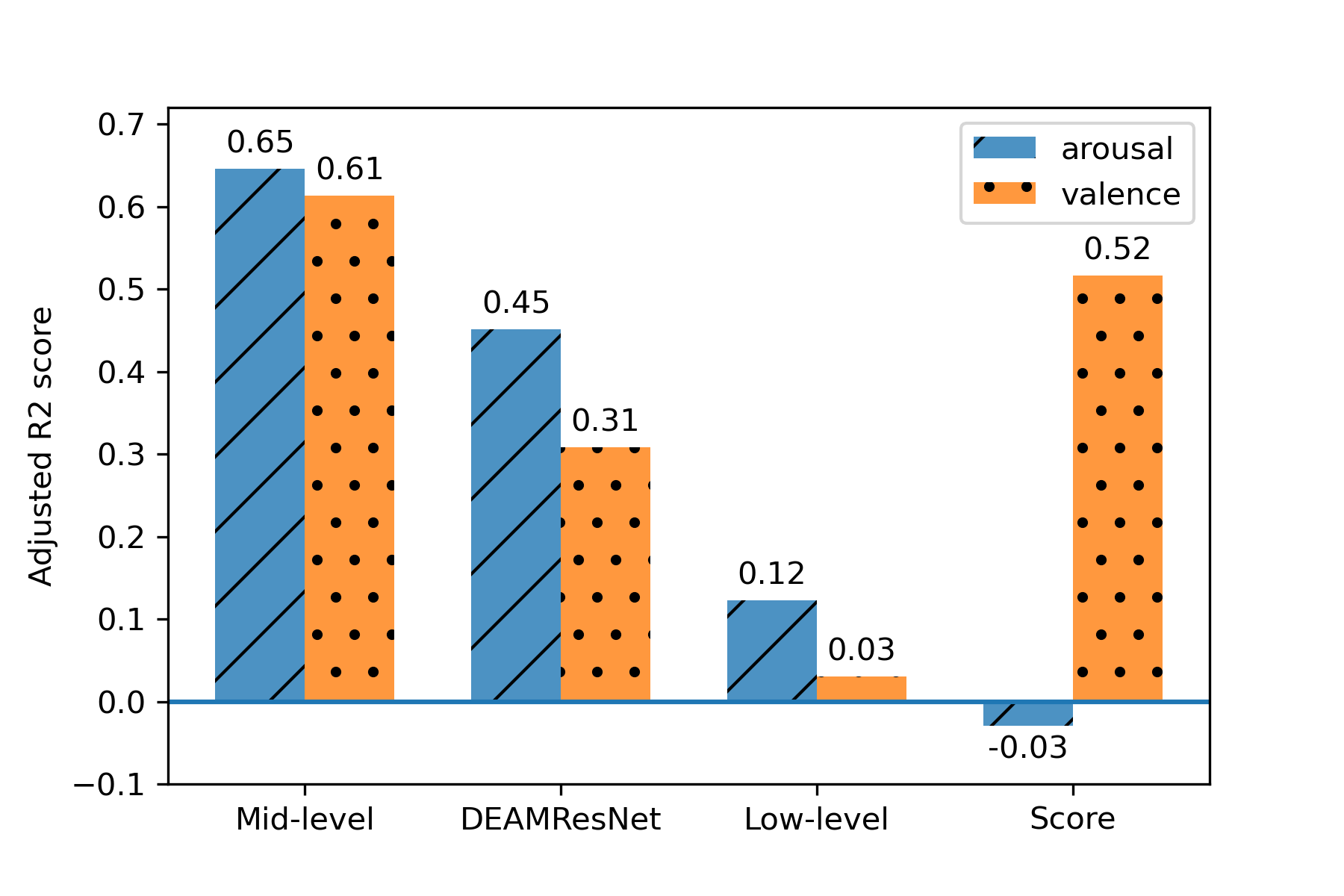}
    \caption{\small $\tilde{R}^2$ scores on outlier performances. The outliers were selected using elliptic envelope on the rated arousal-valence space. Out of 48 pieces, the number of times each pianist was an outlier are Gould: 13, Gulda: 10, Tureck: 10, Schiff: 5, Hewitt: 5, Richter: 5.}
    \label{fig:outlier}
\end{figure}

We therefore create a test set by picking out the outlier performance for each piece in arousal-valence space using the elliptic envelope method~\cite{rousseeuw1999fast}. This gives us a split of 240 training and 48 test samples (the outliers). We train a linear regression model using each of our feature sets and report the performance on the outlier test set in Figure~\ref{fig:outlier}. We see again that Mid-level features outperform the others, for both emotion dimensions. We take this as another piece of evidence for the ability of the mid-level features to capture performance-specific aspects. The surprisingly good performance of score features for valence can be attributed to the fact that for most pieces, the outlier points are separated mostly in the arousal dimension -- the spread of valence is rather small (though not always: see the Gould case in Fig.~\ref{fig:outlier_av})  -- and the score feature ``mode'' is an important predictor of valence (see earlier sections).

\subsection{Predicting Discrete Emotions}
\begin{figure}[h]
    \centering
    \includegraphics[width=\columnwidth]{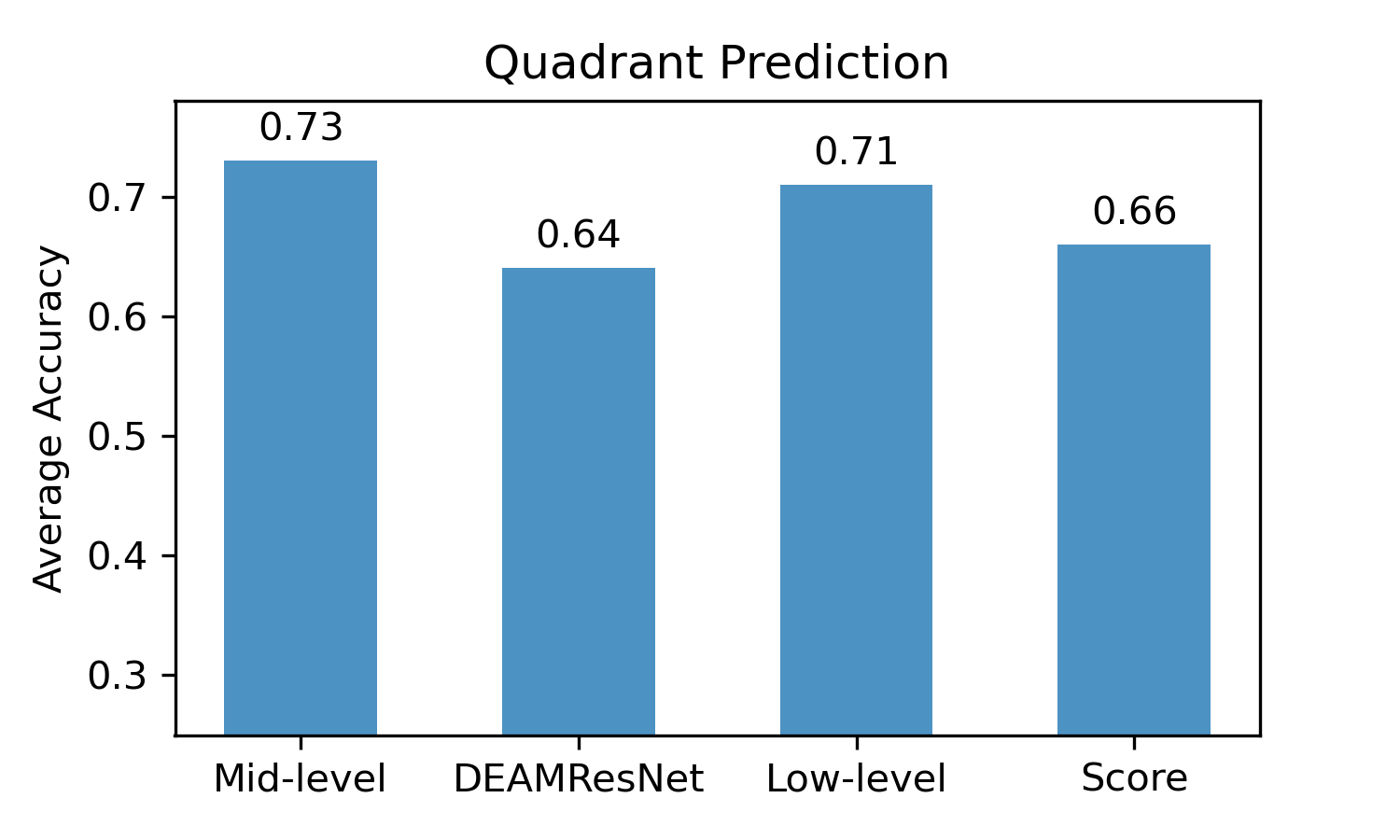}
    \caption{Discrete emotion classification.}
    \label{fig:accuracy}
\end{figure}

Finally, we evaluate how the feature sets perform in a discrete emotion classification task, which might be relevant in a music recommendation setting, for instance. The emotion ratings are converted to classes simply by reducing them to quadrants in the arousal-valence space. In the literature, these are often associated with the basic emotion labels \textit{happy}, \textit{relaxed}, \textit{sad}, and \textit{angry} (in clockwise fashion, starting at upper right). We then train logistic regression models using our feature sets and report the leave-one-out cross-validation accuracy in Figure~\ref{fig:accuracy}. We observe that in this task, all feature sets perform more-or-less equally well, again with a slight advantage for the Mid-level features. Note that the random choice baseline accuracy is 0.25.
An emotion classification model based on mid-level perceptual features might be attractive for performance-emotion-aware music recommendation, being able to offer the mid-level concepts not only as explanations but also as `handles' to search for performances with certain characteristics.

\section{Conclusion}
In this work, we evaluated four feature sets -- mid-level perceptual features, pre-trained emotion features, low-level audio features, and score-based features on their ability to model and predict emotion in terms of arousal and valence. Specific focus was given on the three audio-based features and their modelling power over performance-wise variation of emotion. Mid-level features emerge as the most robust and important among these.

The search for good features to model music emotion is a worthwhile objective since emotional effect is a very fundamental human response to music. Features that provide a better handle on content-based emotion recognition can have a significant impact on applications such as search and recommendation. Modelling emotion is also becoming increasingly relevant in generative music, allowing possibilities such as expressivity- or emotion-based snippet continuation and emotion-aware human-computer collaborative music.

From the experiments presented in this paper, it is clear that deep-learning-based feature extractors are strong competition to the audio features typically used for emotion recognition~\cite{panda2020features}. Here the importance of Mid-level features is even more pronounced -- in addition to being able to model both arousal and valence well under different conditions, they also provide intuitive musical meaning to each feature, and have been previously used as the basis for explainable emotion recognition in~\cite{Chowdhury2019}.

\section{Acknowledgements}
This work is supported by the European Research Council (ERC) under the
EU’s Horizon 2020 research \& innovation programme, grant
agreement No.~670035 (“Con Espressione”), and the Federal State of Upper Austria (LIT AI Lab). The authors would like to thank Carlos Cancino Chacón
for helpful discussions and calculating score-performance alignments and score features, and Jan Schl\"uter and Rainer Kelz for data collection and preparation. Thanks to Aimee Battcock, Cameron Anderson, and Michael Schutz for sharing their annotation data.



\bibliography{ISMIRtemplate}

\end{document}